%% file: main.tex
\documentclass[letterpaper,twocolumn,10pt]{article}
\usepackage{usenix-2020-09}

\usepackage{tikz}
\usepackage{microtype}
\usepackage{graphicx}
\usepackage{subcaption}
\usepackage{booktabs} 
\usepackage{algorithm}
\usepackage{algpseudocode}

\usepackage{hyperref}

\usepackage{filecontents}

\usepackage{titlesec}
\titlespacing*{\section}{0pt}{10pt}{5pt}
\titlespacing*{\subsection}{0pt}{5pt}{5pt}

\usepackage{amsmath}

\newcommand{\sys}{\textsf{WANSpec}}

\newcommand*\circled[1]{\tikz[baseline=(char.base)]{\node[shape=circle,fill=black,text=white,draw,inner sep=1pt,scale=0.75] (char) {#1};}}

\begin{document}

\title{\sys{}{:} Leveraging Global Compute Capacity for LLM Inference}

\author{
{\rm Noah Martin}\\
Tufts University
\and
{\rm Fahad Dogar}\\
Tufts University
} 

\maketitle










\begin{abstract}
Data centers capable of running large language models (LLMs) are spread across the globe. Some have high end GPUs for running the most advanced models (100B+ parameters), and others are only suitable for smaller models (1B parameters). The most capable GPUs are under high demand thanks to the rapidly expanding applications of LLMs. Choosing the right location to run an LLM inference workload can have consequences on the latency of requests due to these high demands. In this work, we explore options to shift some aspects of inference to the under-utilized data centers. We first observe the varying delays affecting inference in AWS services from different regions, demonstrating that load is not spread evenly. We then introduce \sys{}, which offloads part of LLM generation to the under-utilized data centers. In doing so, \sys{} can mitigate capacity issues as well as effectively use on-site compute (ie at universities) to augment cloud providers. This is done with speculative decoding, a widely used technique to speed up auto-regressive decoding, by moving the draft model to the under-utilized compute resources. Our experiments in simulation and cloud deployments show that \sys{} can judiciously employ redundancy to avoid increases in latency while still reducing the forward passes of speculative decoding's draft model in high demand data centers by over 50\%.
\end{abstract}




\input{intro}
\input{background}
\input{motivation}
\input{design}

\input{evaluation}
\input{related_work}
\input{conclusion}

\section{Acknowledgements}
Thanks to the Tufts NAT lab and D.O.C.C. for their feedback and support.

This work was partially funded by NSF CNS award: 2106797.

\bibliographystyle{plain}
\bibliography{main}

\end{document}

%% file: intro.tex
\section{Introduction}
\label{sec:introduction}

The demand for AI is increasing rapidly, with new large language models (LLMs) releasing at a breakneck pace and the demand for GPUs ever growing~\cite{llama4, opus_4_6, kugler2025gpu}. Typically, these models run in data centers with multiple GPUs used for a single inference request~\cite{brown2020languagemodelsfewshotlearners}. As the use cases of LLMs expand, they put increasing pressure on the capacity of GPU compute offered by cloud providers, leading to high prices and limited availability.

\begin{figure*}
\begin{center}
\includegraphics[width=0.8\textwidth]{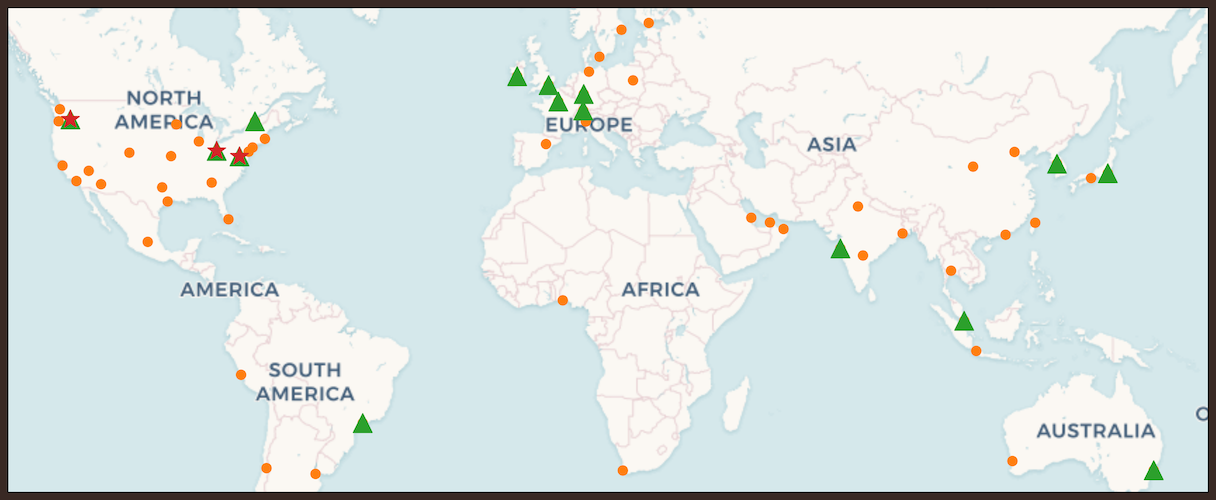}
\caption{Locations of AWS data centers. Stars mark the three regions supporting Claude Opus 4.1 in AWS Bedrock. Triangles mark the more prevalent, but less powerful, Claude Haiku.}
\label{fig:aws_locations}
\end{center}
\end{figure*}

Users of data center compute are given choices in where the models can be run. For example, the AWS Bedrock service is offered in 30 regions distributed around the globe~\cite{bedrock_availability}. However, only a subset of models are available in some regions. While the smaller models (fewer parameters) tend to be offered in more locations. The larger, more capable models tend to be released in limited locations - as shown in Fig~\ref{fig:aws_locations}. The limited availability naturally leads to uneven demand as compute in many regions cannot be used for more capable models. In \S\ref{sec:motivation} we measure this effect as increased tail latency in some regions, to the extent that inference requests for small models could be served faster by using GPUs across continents.

This imbalance in LLM capable data centers raises the question - can we leverage underutilized locations by shifting LLM inference across the wide area network (WAN)? The typical time between tokens for interactive LLM applications is on the order of milliseconds~\cite{grattafiori2024llama3herdmodels}. Whereas data center offloading usually involves non-interactive tasks that may not be reviewed until the next day. At first glance, the difference in time scales makes any amount of WAN shifting seem undesirable for LLM applications. For instance, cloud providers offer cross-region inference as a load balancing solution, but it may increase latency due to the WAN traversal~\cite{cross_region_inference}. Additionally, the heterogeneous compute ability presents yet another obstacle - many regions lack the hardware requirements for the latest model with billions of parameters, and the regions that contain them are not necessarily spread evenly across the globe~\cite{divide-conext}.

The answer to both these challenges is speculative decoding, a technique widely used to speed up LLM inference~\cite{pmlr-v202-leviathan23a}. Speculative decoding employs a smaller draft model to predict tokens which are then validated in parallel by the original target model. This technique speeds up token decoding by as much as 2-3x, and as a result it is widely used in user facing products~\cite{looking_back_speculative_decoding}. The draft model is much smaller, and is a good candidate for running in one of the many other data center locations with less powerful compute. It can also work ahead of the target model asynchronously - generating speculations in parallel with the target model running validations. The overlapping nature of these tasks allows some WAN latency to be absorbed by the pipeline, eliminating its end to end effects.

The main challenge is keeping the decoded tokens in sync across a wide area network without doing too much redundant work or delaying token decoding. To handle this we introduce \sys{}, which judiciously manages token decoding across a WAN, offloading speculative decoding work without compromising on performance. On one side the ``controller'' generates response tokens, and on the other the ``worker'' uses a smaller model to predict multiple steps ahead.

The system takes a full round trip to recover when the worker's prediction is out of sync with the result. During this time the controller falls back to running standard speculative decoding. While the controller runs the target model, the worker is able to catch-up by continuing to run the draft model asynchronously. After this delay, the workers input is again preferred over running on the controller.

To reduce the round trip time stalls, \sys{} predicts when they will happen and if necessary explores multiple output sequences in parallel. This prediction is based on entropy of predicted output tokens - a proxy of model confidence. We run the heuristic on both the controller and worker, each have only partial information of each others state. Tuning the heuristic creates an inherent trade-off between the time to generate a response and amount of offloading.

We evaluate \sys{} using the Llama collection of models~\cite{grattafiori2024llama3herdmodels} and the MTBench~\cite{mtbench} dataset. \sys{} is implemented as a modified version of vLLM~\cite{vllm} deployed in a public cloud environment (\S\ref{subsec:cloud_deployment}). Using data centers in the same georaphic region (US east coast), \sys{} can \textbf{offload 50\% of draft model decoding steps} while \emph{reducing} total generation time compared to standard speculative decoding. We also implement a flexible simulator to test specific aspects of the system under a wide variety of conditions. As we increase the RTT, the benefits of \sys{} gracefully degrade to closely match standard speculative decoding.

The offloading made possible by \sys{} does not need to be in the cloud. For example, a university campus may have an available GPU cluster, or a local/edge device could perform some of the work. The various opportunities for offloading each offer different incentives: cost, load balancing, regulatory compliance, and carbon emission reduction are all reasons one may prefer to shift compute.


Overall, we make the following contributions:

\begin{itemize}
\item Demonstrate uneven load patterns across regions of production cloud inference systems through a global measurement study (\S\ref{sec:motivation}).
\item Design \sys{} to offload inference from costly data centers to excess compute connected over a WAN (\S\ref{design}).
\item Evaluate \sys{} on common LLM workloads using a flexible simulator as well as a cloud data center deployment (\S\ref{sec:evaluation}).
\end{itemize}

%% file: background.tex
\section{Background}
\label{sec:background}

In this section we provide an overview of speculative decoding and some of it's variations.


\subsection{LLM Inference}
Inference of large language models (LLMs) is composed of two stages - a prefill and decode. The prefill processes the input tokens in parallel and is compute intensive. The decode stage is memory intensive and takes the majority of the time, due to the auto-regressive nature of LLMs. This stage generates output tokens step by step, with each forward pass determining a new token that depends on the previous ones.

The prefill phase creates the KV cache - a key enabler of efficient LLM inference that prevents recomputing the key and value vectors used by the attention mechanism. The KV cache grows linearly as each token is generated in the decode stage. Effectively managing the large amount of memory used by this cache makes high throughput serving possible in LLM serving systems such as PagedAttenion~\cite{vllm}, LServe~\cite{yang2025lserve}, and FlashAttention~\cite{dao2022flashattentionfastmemoryefficientexact}.

\subsection{Speculative Decoding}
\label{sec:speculative-background}

Speculative decoding~\cite{pmlr-v202-leviathan23a} and speculative sampling~\cite{chen2023acceleratinglargelanguagemodel} are techniques, inspired by speculative execution in a CPU, to speed up auto-regressive LLM decoding while leaving output distributions unchanged. They achieve speedups by parallelizing token generations using a \emph{draft model}. The draft model generates \emph{candidate tokens} which are then validated in parallel by the \emph{target model}. Using this technique LLM token generation can be 2-3x faster.

The draft model is typically an order of magnitude smaller than the target model, and runs much faster. After $k$ iterations of the draft model, one iteration of the target model can generate the next token for $k$ sequences in a single batch. In the best case, all candidate tokens were validated and $k$ iterations of the draft model plus 1 iteration of the target model produces $k+1$ output tokens.

Due to the large performance gains, speculative decoding has been applied throughout the industry, even in Google’s flagship products~\cite{looking_back_speculative_decoding}. In our work, we take speculative decoding and apply it across a wide area network. This creates flexibility in datacenter load, which can reduce costs, without sacrificing on performance.

Many variations and improvements on speculative decoding have been proposed. For instance, some increase the draft acceptance rate by building up a token tree and validating multiple branches in parallel~\cite{opt-tree, spec_infer} or by improving candidate token accuracy through custom training procedures~\cite{cai2024medusasimplellminference, li2024eagle2fasterinferencelanguage}. Others parallelize the draft and target models to realize further improvements~\cite{zhang2025swiftspecultralowlatencyllm}.

Prior work on speculative decoding does not consider the wide distribution of devices capable of LLM compute, and opportunities for connecting them over a WAN. As described in~\S\ref{sec:introduction}, many of these locations are only suitable for smaller models, which play a critical role in speculative decoding. Based on this observation, \sys{} can judiciously employ the additional compute to reduce load and cost budgets while maintaining the performance benefits of speculative decoding.

%% file: motivation.tex
\section{Motivation}
\label{sec:motivation}


\begin{figure*}
    \centering
    \subfloat[][Median]{\includegraphics[width=0.45\textwidth]{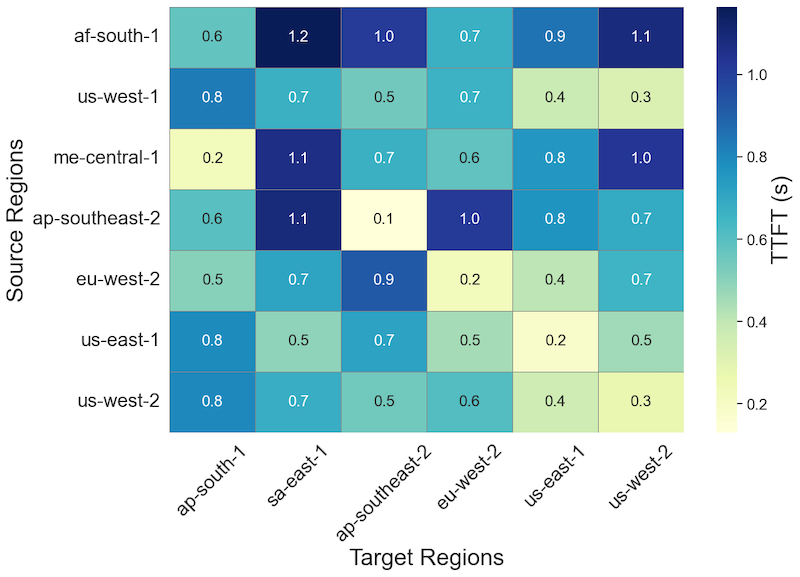}
    \label{subfig:p50_heatmap}}
    \subfloat[][p95]{\includegraphics[width=0.45\textwidth]{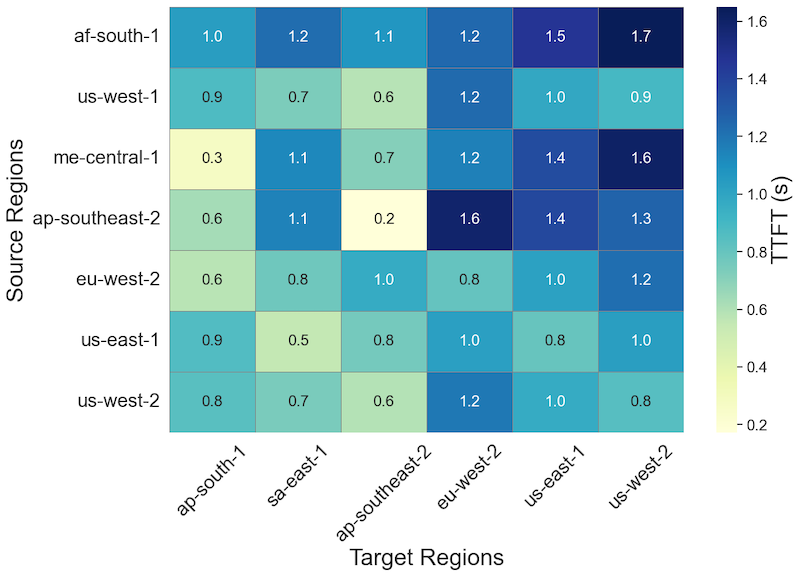}
    \label{subfig:p95_heatmap}}

    \caption{Median (p50) and tail (p95) time to first token of Claude 3 Haiku between AWS regions. Each cell shows p50/p95 latency (in seconds) for a request originating from the source region (rows) to a target Bedrock region (columns). The data were collected over 3 days. In~\ref{subfig:p50_heatmap} intra-region latencies are lowest, as expected. In~\ref{subfig:p95_heatmap}, some regions---particularly \texttt{eu-west-2}, \texttt{us-east-1}, and \texttt{us-west-2}---exhibit the lowest latencies for inter-region requests. This suggests inference queuing dominates network latency in theses cases.}
    \label{fig:heatmap_results}
\end{figure*}

The success of LLMs across a wide variety of applications has led to a rapid rise in their usage and increasing complexity of models. Together, these drive up the demand for their underlying compute resource: GPUs. Data centers using these GPUs are now a critical resource to run LLM applications as they require the latest generation of advanced hardware~\cite{dynamollm}. The latest generation GPUs typically become available in a small subset of cloud regions, making them also a scarce resource.

The high demand for this infrastructure is likely to create delays such as queuing and rate limiting. AWS support documentation even states that inference requests may be restricted during peak usage, and suggests mitigating this by performing inference in different regions~\cite{cross_region_inference}. However, when these regions lack the available hardware to meet the needs of the largest models they may go under-utilized. Beyond the cloud, other sources of GPU compute such as home, office, or edge deployments can be available to users of cloud compute but go unused. This motivates us to shift some inference work required for the latest models to compute across the network.

To understand how LLM load is distributed in cloud data centers we perform a measurement study of inference latency across different AWS regions. We speculate that some popular regions are in higher demand than others, creating an uneven geographic distribution of inference latency caused by queuing. While previous measurement studies~\cite{wang2025burstgptrealworldworkloaddataset}
 have shown periodicity in LLM usage focused on a single region, our study highlights the variation across regions. We find some regions experience consistently higher delays than others, and some show diurnal patterns. The delay can even be high enough that lower end to end latency is achieved with a region in another continent - the larger network latency is made up for by reduced delay in the data center.


Our measurement study was run in 7 AWS regions (the source regions) across the globe, over 3 days, and targeted LLM infrastructure in 6 regions (the target regions). Each measurement is the time to first token (TTFT) of the Claude 3 Haiku model provided by the target region’s AWS bedrock service. The input is less than 10 tokens to minimize prefill time and focus on other sources of delay. Every 15 minutes the measurement was run 40 times. We first ran the experiment in August 2025 and repeated it in January 2026 to confirm the results are still applicable. The second experiment also included TCP connect time, to rule out changes in network latency as the cause of delay. In total over 900k samples were collected.

The heatmaps in Fig~\ref{subfig:p50_heatmap} show distance from source to destination region generally correlates to median response time for a single token generation. As expected, TTFT is low for the 4 setups that run the same source and target region where network latency is minimal. However, Fig~\ref{subfig:p95_heatmap} paints a very different picture of tail TTFT. In this heatmap, LLMs in the target regions \texttt{eu-west-2}, \texttt{us-east-1}, and \texttt{and us-west-2} exhibit much higher tail TTFT than the other regions - even for intra-region requests. In fact, for requests originating in these 3 regions the p95 TTFT was lowest when making an inter-region request to a different continent. This indicates queuing delay is higher than the network RTT across continents.

\begin{figure}
\begin{center}
\includegraphics[width=\columnwidth]{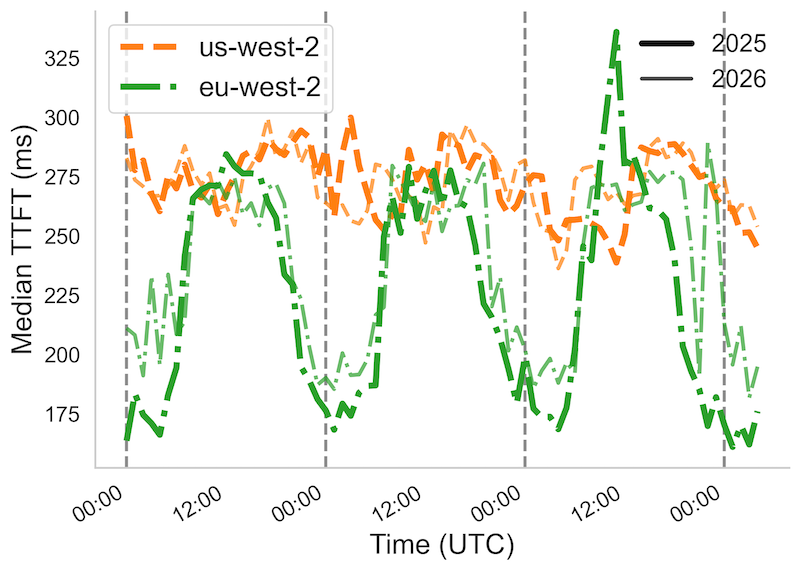}
\caption{Results of AWS Bedrock measurements over three days, for the same source and target region. Some regions (\texttt{eu-west-2} in this graph) exhibit diurnal patterns, while others (\texttt{us-west-2}) do not. This pattern held in our original experiment in 2025 as well as the 2026 repeat.}
\label{fig:diurnal-ttft}
\end{center}
\end{figure}

To investigate this further we looked for patterns in how TTFT changes throughout the day. The results in Fig~\ref{fig:diurnal-ttft} show clear diurnal patterns for median latencies in \texttt{eu-west-2}, with higher latencies during the local daytime. This is expected if increased load during the day causes queuing or contention. While a few regions exhibited this trend for their local daytime, others such as \texttt{us-west-2} in Fig~\ref{fig:diurnal-ttft} do not. These trends throughout the day held for our 2025 and 2026 experiment.

\begin{figure*}
    \centering
    \subfloat[][Median TTFT]{\includegraphics[width=0.3\textwidth]{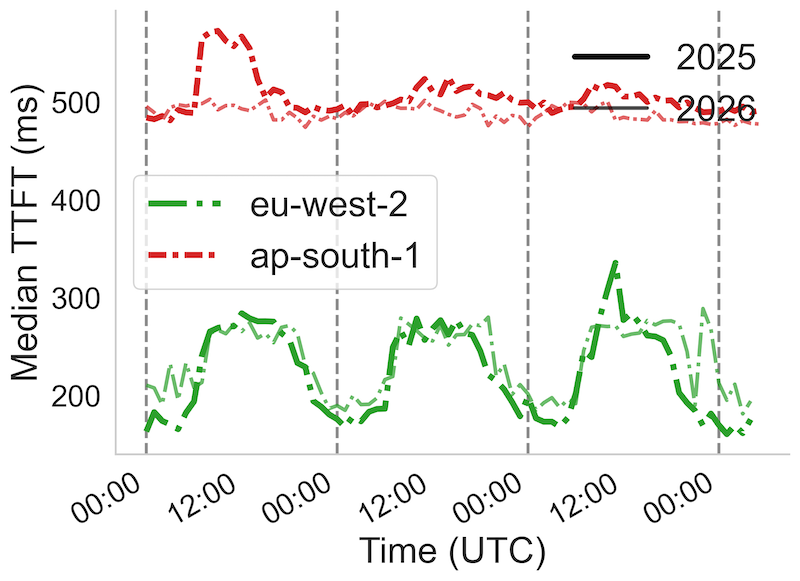}
    \label{subfig:eu-p50}}
    \subfloat[][p90 TTFT]{\includegraphics[width=0.3\textwidth]{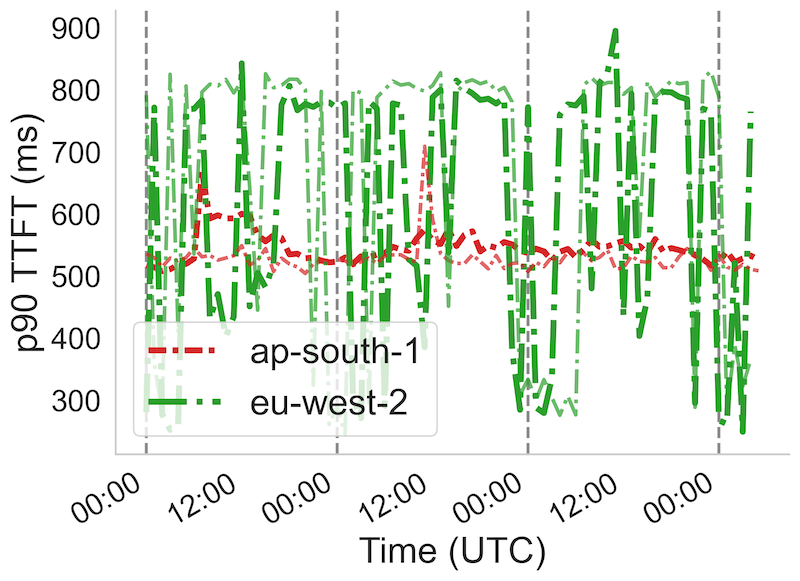}
    \label{subfig:eu-p90}}
     \subfloat[][TCP connect]{\includegraphics[width=0.3\textwidth]{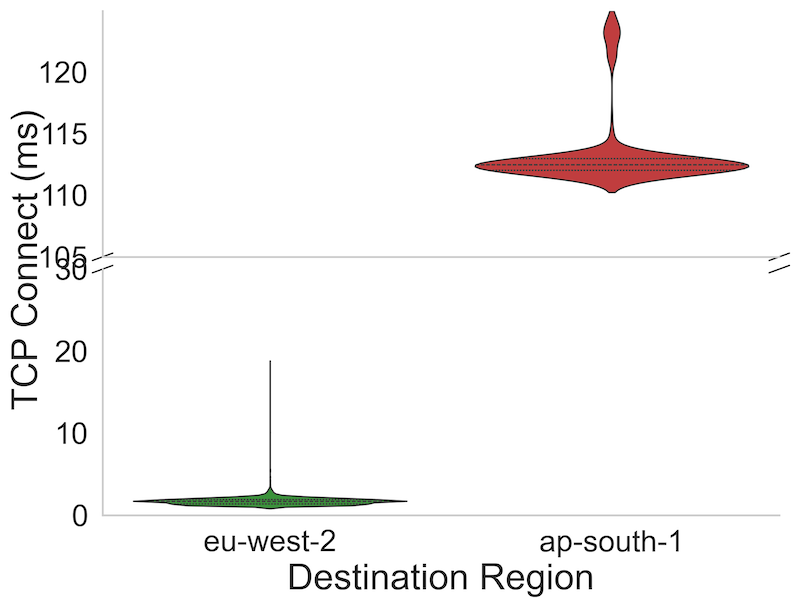}
    \label{subfig:network-latency}}

    \caption{Results of our experiments with requests originating in \texttt{eu-west-2} and targeting \texttt{eu-west-2} and \texttt{ap-south-1}. \ref{subfig:eu-p50} and \ref{subfig:eu-p90} plot the median/p90 TTFTs for each hour. The median intra-region requests exhibit diurnal patterns. At p90, the intra-region TTFT shows significant instability, while the inter-region requests are more stable and at times lower latency. To help rule our network latency as the bottleneck, Fig~\ref{subfig:network-latency} plots the distribution of TCP connects we measured during our 2026 experiment, with the top 1\% of outliers excluded for readability. Intra-region requests have low latency. Requests across the WAN have higher, but still stable, latency. The stability of the latency but instability of TTFT indicates increased queuing causes the slowdown.}
    \label{fig:measurement_experiment_results}
\end{figure*}

Fig~\ref{subfig:eu-p50} and Fig~\ref{subfig:eu-p90} zoom in on TTFT for requests originating in \texttt{eu-west-2} and targeting another continent as well as itself (intra-region). The graphs show median and p90 for each hour of data, respectively. As expected, the median latencies are lowest when the LLM is running in the same region the request is sent from - due to minimal network delay. Even the peak during the diurnal patterns are lower than the p50 TTFT of a cross-continent request. However, the tail results are very different. There are large spikes in TTFT for \texttt{eu-west-2}. In contrast, \texttt{ap-south-1}, the geographically furthest region, has relatively consistent p90 TTFT. At some times, the time to generate a token from the region on a different content is lower than the intra-region request.

Measurements from our vantage point are limited, because we must treat AWS bedrock as a black box. The most likely cause of the observed change in inference time is that some regions have higher GPU load than others, creating queuing delay. Specifically, \texttt{ap-south-1} appears to suffer less from excessive queuing delay in Fig~\ref{fig:measurement_experiment_results}, and \texttt{eu-west-2}, \texttt{us-east-1}, \texttt{us-west-2} all have higher queuing delay in Fig~\ref{fig:heatmap_results}. Other factors such as higher tail network delay on some paths could also cause this result. However, this is less likely due to the reliable infrastructure in private cloud networks~\cite{judicious_qos}. To confirm this, we also measured TCP connect latency as a proxy for network delay during the 2026 experiment. Fig~\ref{subfig:network-latency} shows this measured latency for our intra and inter region experiments in Fig~\ref{fig:measurement_experiment_results}. The network delays are shown to be consistent, and no large fluctuations account for the changes in TTFT. The intra-region requests, which do not traverse a WAN, have the consistent low latency as we expect.

These results motivate the need to distribute computation over available GPUs -- even if they are in different regions. Speculative decoding, with its smaller draft model, is amenable to inference-time load shifting as long as the challenges of WAN latency can be overcome.

%% file: design.tex
\section{WANSpec}
\label{design}

We now describe the design of \sys{}, a system that accelerates inference by leveraging available GPU resources across wide-area networks (WANs). The core idea behind \sys{} is to judiciously use redundant computation to hedge against the additional latency incurred when a WAN traversal is expected. \sys{} generates responses using speculative decoding, with some draft model token generation offloaded to a WAN-connected machine. At each side of the network, components of \sys{} predict when the candidate tokens will not match with the target model and perform extra token decoding to mitigate the mismatch.

\subsection{Challenges and Opportunities}
\label{subsec:challenges}

The main challenge that our system must overcome is to keep the sequences generated by draft and target models in sync with a potentially high RTT between them. It will take at least one RTT for the target model to correct the draft and start receiving updated speculations. In typical speculative decoding, this delay is negligible because the same GPUs are used for both target and draft models, or they are run in parallel on GPUs connected to a single machine~\cite{zhang2025swiftspecultralowlatencyllm}.

When the draft model predicts a token that doesn't match the target, a synchronization step is necessary. This step involves sharing the correct token from the target model and then receiving the first updated candidate token. This synchronization takes one RTT plus the time to generate a draft token, and until it is completed the candidate tokens cannot be used. For large RTTs this creates a significant stall, that would slow speculative decoding to a point of diminishing returns even if the draft model only infrequently differs from the target.

Fortunately, this distributed setup also poses some opportunity which \sys{} exploits to overcome the challenges. Running both target and draft models in parallel creates an inherent \textbf{slack} for draft token generation. As long as the draft model is started sometime before the target model completes, the process does not take any longer than speculative decoding described in~\S\ref{sec:speculative-background}.

\begin{figure}
\begin{center}
\includegraphics[width=\columnwidth]{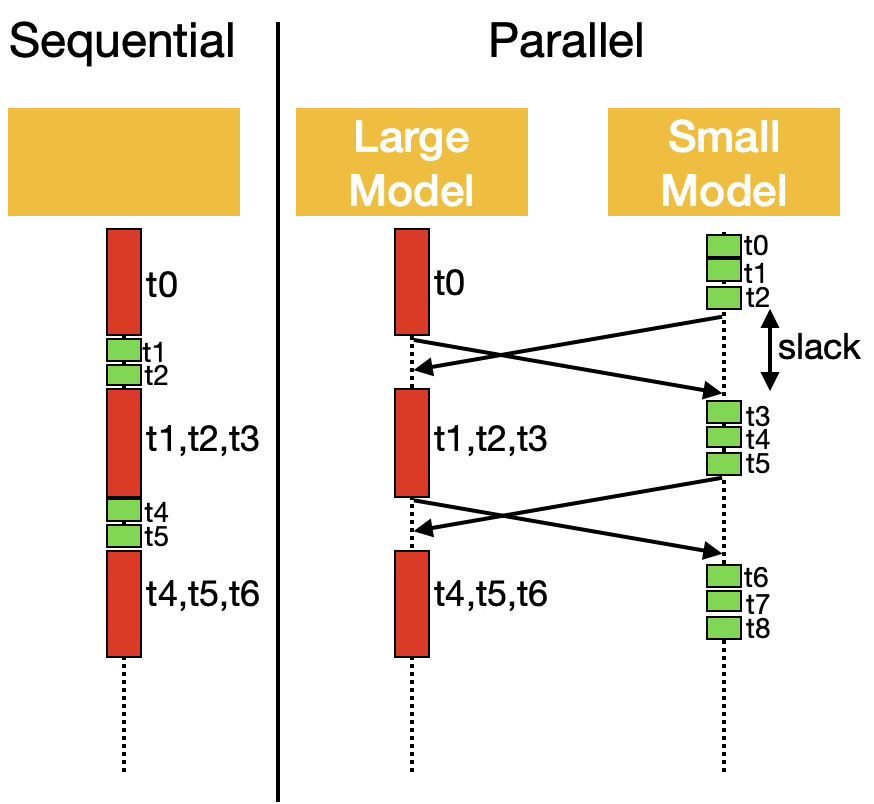}
\caption{Comparison of using speculative generation sequentially vs. in parallel. Overlapping draft and target model forward passes creates slack which can mask network latency. This parallelization requires one extra forward pass of the draft model per token generated by the target model.}
\label{fig:speculative-slack}
\end{center}
\end{figure}

Fig~\ref{fig:speculative-slack} illustrates this slack time and how it can be used to mask RTT between the two models. In the figure, red blocks represent the time to run the target model, and green blocks are the time to run the quicker draft model. On the sequential side, target and draft model computation do not overlap and therefore draft token generation is on the critical path. On the parallel side, the draft model runs simultaneously with the target. As long as draft model tokens are delivered over the WAN before the next iteration of the target model is scheduled the total time remains the same.

However, there are two drawbacks to this approach that can cause latency to increase if the RTT is not appropriately managed. The first is that an extra draft token must be generated sequentially for each step of the target model. In Fig~\ref{fig:speculative-slack} these are t0, t3, and t6 which are part of the draft sequence. Additionally, a mismatch between draft and target models would prevent masking of the slack. In the Fig~\ref{fig:speculative-slack} example, the draft model begins work on t4 before the target has computed t3. If the draft t3 and target t3 do not match, the t4 computation would need to be restarted, stalling generation by one round trip.

\subsection{Design Overview}

\sys{} avoids the round trip time delay by judiciously adding redundant generation when it helps to avoid a stall in token generation. The challenges described in~\ref{subsec:challenges} stem from a disagreement between draft and target models, but at each side of the network only one half the information is available to tell if a disagreement has occurred. However, the inherent probabilistic output of LLMs gives us a proxy of the model's ``confidence'', which \sys{} uses to predict a mismatch. \sys{} is composed of two pieces that work together to implement this solution. The components are the \textbf{controller}, which uses its greater compute resources to run the target model, and the \textbf{worker}, which only runs the draft model. Each of these components \emph{predict} when a synchronization is likely and take steps to mitigate the effect through redundant computation. The worker does this by exploring multiple candidate sequences in parallel, and supplying the alternative sequences to the controller so they are available if the first candidate cannot be used. The controller does this by predicting when the next useful token will arrive from the worker, and runs the draft model directly if latency is expected to exceed the time to generate the next token locally.

The prediction on each component must use opposite pieces of information. The controller knows the target model output but not the draft model output, and vice versa on the worker. In both cases, \sys{} employs the prediction entropy, of the target model on controller and draft model on worker, as a measure of model confidence which determines if extra generation steps need to be added. This is inspired by prior work which has used entropy as a probabilistic measure of confidence~\cite{edge_bert}. We explain the details of this process on the controller in \S\ref{sec:design-controller}, and on the worker in \S\ref{sec:design-worker}.

\begin{figure}
\begin{center}
\includegraphics[width=\columnwidth]{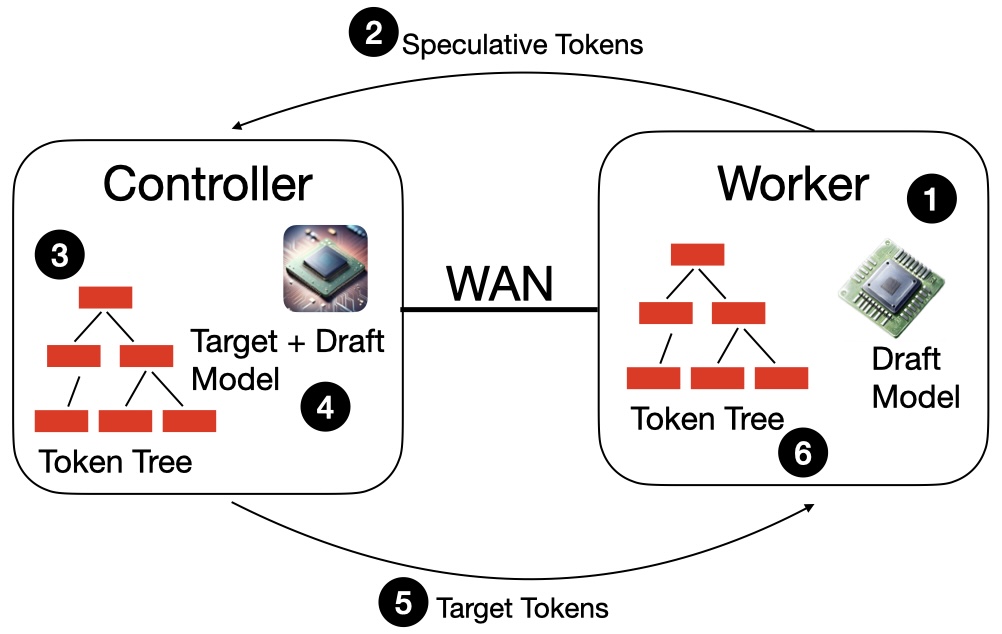}
\caption{\sys{} splits speculative generation across a wide area network (WAN) to offload inference work to underutilized compute resources. The controller runs the target model, and optionally the draft model, while the worker runs only the draft model to provide speculations to the controller. Both components judiciously add extra draft model decode steps to mask the WAN latency.}
\label{fig:design-overview}
\end{center}
\end{figure}

Fig~\ref{fig:design-overview} shows how these pieces work together at a high level. The worker \circled{1} can only run the draft model and streams tokens over a WAN \circled{2} to the controller which stores the tokens that match an already generated prefix \circled{3}. The controller can run the draft model to generate more candidate tokens, and once $k$ are ready the controller runs the target model \circled{4}. Finally, the controller streams tokens back over the WAN to the worker \circled{5} where they are used to prune the speculations \circled{6}. Since the worker only runs the draft model, it does not need as many resources as the controller. For instance - it could be in a datacenter with fewer GPUs, or an older generation of hardware. It is the worker's job to generate the most useful speculations with the draft model, and the controller's job to keep the worker up to date while not delaying steps of the target model.

\subsection{Controller}
\label{sec:design-controller}
On one side of the WAN, the controller runs the target model once $k$ candidate tokens are available. The resulting target model token predictions are sent across the network to the worker. The controller maintains its own speculative tree, added to when new speculations are received from the worker and pruned after each step of the target model. The controller must also estimate when the next usable token will arrive from the worker, using the timestamp of the last update and the expected round trip time ($R$). When the next token is estimated to take longer to arrive than the time to run the draft model, the controller will switch to running the draft model to avoid a stall. Algorithm~\ref{alg:controller} shows the process run on the controller.

At each large model step $1$ to $k+1$ tokens are generated. If less than $k+1$, the worker is marked out of date so the controller can switch to running the draft model until a new correct speculation is received. When $k+1$ are generated there is not enough information at the controller to know if the last token will match the next speculation. We use entropy of the target model’s last generated token to estimate the match likelihood. If the entropy is above a threshold, $\phi$, the controller assumes the draft model needs updating. If this is wrong, the controller may do extra work or the generation may unnecessarily stall while no useful speculations are generated. This tradeoff based on the value of $\phi$ is explored further in \S\ref{sec:evaluation}.

\begin{algorithm}[tb]
   \caption{Controller}
   \label{alg:controller}
\begin{algorithmic}
   \State Initialize $token\_tree, t_{update}$
   \State Input $k, R, T_{draft}, \phi$
   \Repeat
   \State $token\_tree.append(receive\_speculations())$
   \If{$token\_tree.depth >= k$}
   \State $result = step\_target\_model()$
   \State $send(result)$
   \State $token\_tree.update(result)$
   \If{$result.length < k+1$}
   \State $t_{update} = current\_time$
   \ElsIf{$result.entropy > \phi$}
   \State $t_{update} = current\_time$
   \EndIf
   \ElsIf{$t_{update} + R > current\_time$}
   \State $token\_tree.update(step\_draft\_model())$
   \EndIf
   \Until{$EOS$}
\end{algorithmic}
\end{algorithm}

\subsection{Worker}
\label{sec:design-worker}
The worker is responsible for running the draft model and providing candidate tokens to the controller. The more accurate the candidate tokens are, the faster decoding can proceed. The worker must avoid going too far down a wrong path, generating a sequence that will be discarded by the controller. It does this by judiciously generating extra tokens in alternate sequences.

Since the draft model may get far ahead of the target, particularly when RTT is high, more than $k$ speculations could be decoded before they reach the controller. In this case the worker does not need to spend all its time generating a single sequence. Instead, to reduce mismatches, the worker can consider multiple candidate tokens at each step and construct a token tree. Each path from root to leaf of the tree represents a sequence of candidate tokens. Trees such as these have been used for speculative decoding to increase its speedup~\cite{spec_infer}. We use them to reduce the chance that the worker's token will be discarded after waiting for a synchronization with the controller.

The worker cannot explore all possible sequences at each generation step, the size of this tree would grow exponentially and is only pruned when the worker receives new tokens from the controller. Instead, \sys{} must decide which parts of the tree to prioritize. We use a fixed branching factor, $b$, and predict the need to perform each possible branch. Once again, this prediction is based on token entropy, this time of the draft model. When the entropy of a token is below a threshold, $\theta$, the branching is skipped. This follows from our intuition that a low entropy token demonstrates confidence in the prediction and is less likely to require correction by the target model. Prior work has used beam search for similar purposes, and this could be used in \sys{} as well~\cite{li2024eagle2fasterinferencelanguage}. We plan to explore this and other alternative strategies for constructing the tree in future work.

Details of the worker is provided in Algorithm~\ref{alg:worker}. The parameter $s$ is the maximum number of parallel sequences that can be considered in one batch and limits how wide the speculative tree can grow. The choice of this parameter will depend on hardware constraints of the system.

\begin{algorithm}[tb]
   \caption{Worker}
   \label{alg:worker}
\begin{algorithmic}
   \State Initialize $token\_tree$
   \State Input $b, \theta, s$
   \Repeat
   \State $token\_tree.prune(receive\_validations())$
   \State $candidates = token\_tree.most_probable(s)$
   \State $output = step\_draft\_model(candidates)$
   \For{$p$ in $ouput$}
   \If{$p.entropy < \theta$}
   \State $results = argmax(p)$
   \Else
   \State $results$ = $(argmax(p), argmax_2(p))$
   \EndIf
   \State $send(results)$
   \State $token\_tree.update(results)$
   \EndFor
   \Until{$EOS$}
\end{algorithmic}
\end{algorithm}

\subsection{KV Cache Consistency}

The worker and controller must both maintain a draft model KV cache. While the worker generates each token auto-regressively, the controller only generates a draft model token when it predicts a synchronization is necessary. At this point the controller may have multiple known tokens generated by the worker. Rather than evaluating the draft model sequentially, the controller executes a batch request of all known tokens that have not been handled yet - essentially running a prefill. In our experiments (\S\ref{subsec:cloud_deployment}), this does not increase overall time. However, if the controller's target model gets too far ahead of the draft the batch size can exceed available GPU memory. We impose a limit on the size of the batch and execute multiple batches sequentially if necessary.

\sys{} may also sync the KV cache from worker to controller over the WAN, but this increases the amount of data transfer. In future work we plan to explore various optimizations to improve this, such as quantization or cache compression~\cite{kvpress}. Lossy compression will not impact the accuracy of \sys{} because this is the draft model KV cache, not the target model. Our implementation in \S\ref{sec:evaluation} does not transfer the KV cache, and only computes it in parallel as described previously.



%% file: evaluation.tex
\section{Evaluation}
\label{sec:evaluation}

We evaluate \sys{} using a flexible simulator to explore the effects of our optimizations (\S\ref{subsec:ablation}, \S\ref{subsec:latency_v_tokens}), and a cloud deployment using vLLM (\S\ref{subsec:cloud_deployment}).

\subsection{Setup}
\label{subsec:setup}
Our event based simulator is implemented in Python and explore the benefits of \sys{} under different conditions. Time to run each step of the target and draft models are taken from the reported time to run Qwen2-72B and Qwen2-1.5B on 8 NVIDIA 80GB H800 SXM GPUs~\cite{zhang2025swiftspecultralowlatencyllm}. The simulations make non-overlapping sequential requests and generate 100 tokens for each request, by drawing randomly without replacement from a dataset of 1491 outputs taken from generating responses to the MTBench~\cite{mtbench} questions with Llama-3.3-70B and Llama-3.2-1B~\cite{grattafiori2024llama3herdmodels}. The match rate of these tokens is $\sim$80\%. Our method assumes tokens are i.i.d., consistent with how speculative decoding has been evaluated in prior work~\cite{chen2023acceleratinglargelanguagemodel, pmlr-v202-leviathan23a}.
We use a greedy sampling strategy at each decoding step of the target and draft models.

\subsection{Ablation study}
\label{subsec:ablation}

To demonstrate the effects of \sys{} we progressively enable each optimization of the simulator and compare it with standard speculative decoding. We start by adding a branch factor of 2, then the worker entropy threshold, $\theta=0.5$, and finally the controller entropy threshold, $\phi=0.5$. We tested a range of round trip times, and for each system configuration take the median of 20 iterations.

\begin{figure}
    \centering
    \subfloat[][Latency]{\includegraphics[width=0.48\columnwidth]{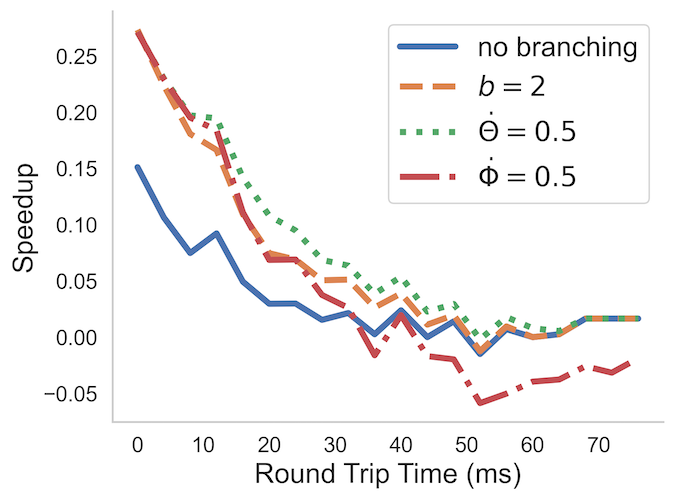}\label{subfig:speedup}}
    \subfloat[][Tokens]{\includegraphics[width=0.48\columnwidth]{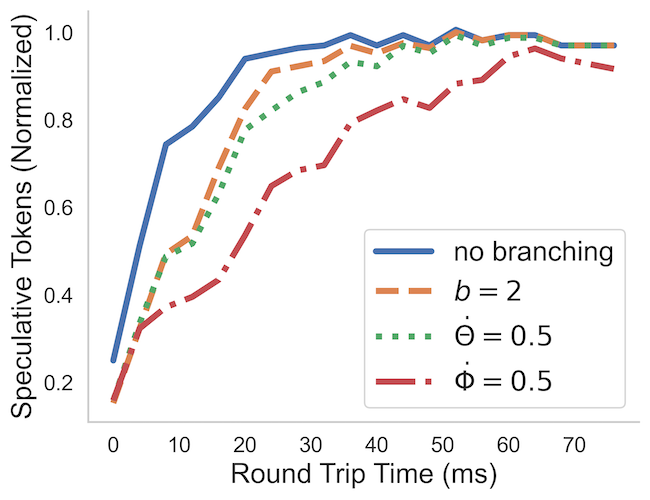}\label{subfig:tokens}}

    \caption{Median latency (\ref{subfig:speedup}) and draft tokens generated by the controller (\ref{subfig:tokens}) with various settings in \sys{}. Both are relative to the result of running standard speculative generation on just the controller. Enabling an entropy threshold on the controller negatively impacts the latency to generate responses, but reduces the number of draft model forward passes.}
    \label{fig:microbenchmarks}
\end{figure}

The results in Fig~\ref{subfig:speedup} shows the improvements in latency to generate a response. Without any branching factor the speedup over speculative decoding is nearly 0 when the round trip time is only 10ms. The use of a speculative tree with 2 branches is the optimization that makes the most difference, the RTT that still provides a benefit nearly doubles to around 20ms. When we enable the worker entropy heuristic to prune the speculative tree, it is able to spend more time on the most likely speculative sequences, achieving more performance benefits. The speedup with a 20ms RTT approximately doubles to a 10\% improvement over speculative decoding. The second entropy heuristic, on the controller, decreases the performance gains because it is used to optimistically skip an extra forward pass of the draft model. When this heuristic is wrong, \sys{} makes no progress until it is corrected by synchronizing across the WAN. For longer RTTs this can cause \sys{} to generate slower than speculative decoding, but the slowdowns in this example are under 5\%.

While the speedup (and in some cases slowdown) from \sys{} is modest, the primary benefit comes from the reduced draft model forward passes performed by the controller shown in  Fig~\ref{subfig:tokens}. With no optimizations \sys{} cannot reduce the draft model passes except in the lowest RTT settings. The introduction of branching, and the worker's entropy threshold, make some improvements in the 10-15ms range, but the largest benefit comes from the entropy optimization on the controller. This is what allows \sys{} to skip running the draft model when a token is expected to be received from the worker. With these optimizations, \sys{} reduces the controller draft model forward passes by 50-30\% in the 20-30ms RTT range. This is within the expected latency of data center regions on the same continent such as AWS \texttt{us-east-1} and \texttt{ca-central-1}.


\begin{figure}
    \centering
    \includegraphics[width=0.98\columnwidth]{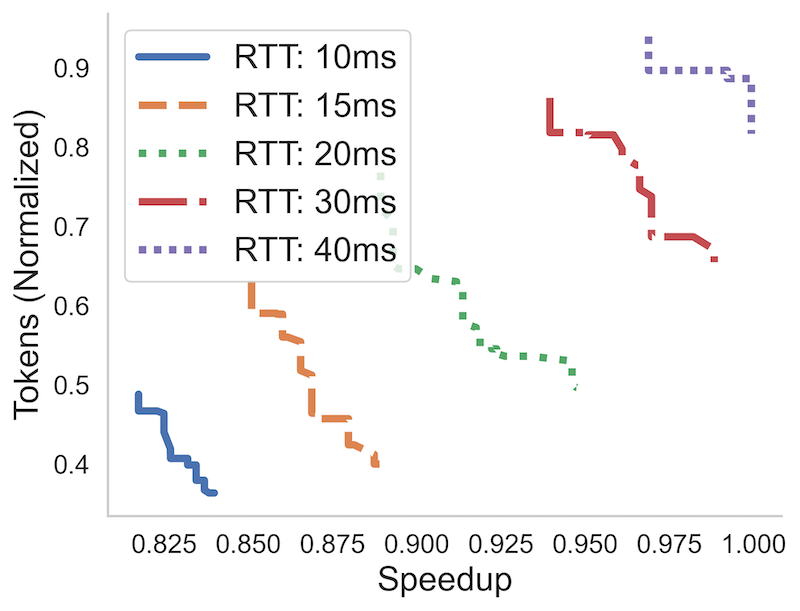}

    \caption{Tradeoff between time to generate responses and steps of the draft model taken by the controller. Each line represents a different round trip time and a sweep of the entropy threshold: $\phi$. Using the entropy threshold allows exploring this trade off in a nearly Pareto optimal manner.}
    \label{fig:tradeoff}
\end{figure}

\subsection{Latency vs. Tokens Generated}
\label{subsec:latency_v_tokens}

The ablation study shows how our four optimizations enable \sys{} to deliver benefits even in WANs with higher latency. Most optimizations were purely beneficial; however, the value of $\phi$ creates a trade off. When it is lower, the controller will more often generate a draft token. This prevents a stall if the worker's prediction is incorrect, but comes at the cost of more load on the controller. Consider the following example sequence of events on the controller:

\begin{enumerate}
\item The controller validates both candidate tokens and decodes 3 new output tokens.
\item The controller sends the 3 newly generated tokens to the worker across the WAN.
\item The controller waits for the next candidate token, expecting it to arrive shortly and match the 3rd just generated token.
\item The next candidate token arrives, and it does not match the 3rd output token generated in step 1.
\item The controller starts generating a new candidate token, what the worker sends will be discarded until one RTT after step 2.
\end{enumerate}

In this example, it would have been more optimal to have the controller start generating a token at step 3 rather than waiting. However, if the next received token had been correct, it would have barely provided a speedup at the expense of an extra draft model step.

In Fig~\ref{fig:tradeoff} we show the latency-load trade off for various values of $\phi$ at different round trip times. Each curve is a configuration with a specific RTT, and sweeps 100 values of $\phi$ spanning the smallest to largest entropy observed in the dataset. Results indicate our entropy optimization is a good way to navigate this tradeoff space, with most points lying on a Pareto-optimal curve.

The change in latency from choosing different values of $\phi$ is not as high as the affect on candidate tokens generated by the controller. In most cases the latency difference is under 5\% of the time for standard speculative decoding. Due to this, most applications would benefit from a balanced approach, benefiting from large reductions in extra draft model passes while decreasing speed only slightly.

Based on the results in Fig~\ref{fig:tradeoff}, \sys{} can reduce draft tokens generated by over 30\% up to 30ms RTT. By tuning $\phi$ we can even achieve 20\% reductions at 40ms RTT. These are within the expected RTT of many cloud datacenter regions located on the same continent, such as regions throughout Europe. Additionally, for users who benefit from having a nearby cloud data center, it is within expected latency of their local/onsite compute~\cite{divide-conext}. This provides opportunities for further cost reductions by using GPUs available locally for some compute instead of more capable and expensive cloud resources. However, this latency is not within a range that would allow load shifting based on diurnal patterns, since the timezone differences required to take advantage of this would be across continents.

\subsection{Cloud Deployment}\label{subsec:cloud_deployment}

\begin{figure}
    \centering
    \subfloat[][Latency]{\includegraphics[width=0.48\columnwidth]{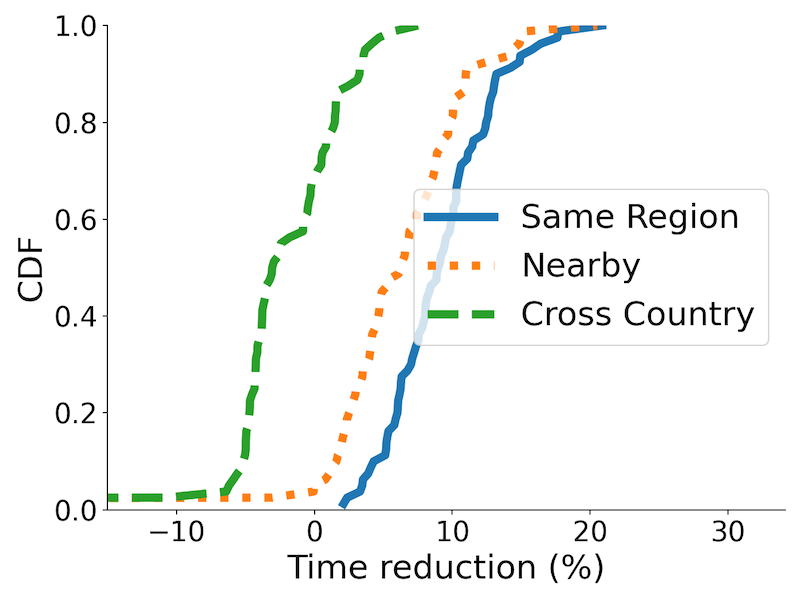}\label{subfig:time-vllm}}
    \subfloat[][Tokens]{\includegraphics[width=0.48\columnwidth]{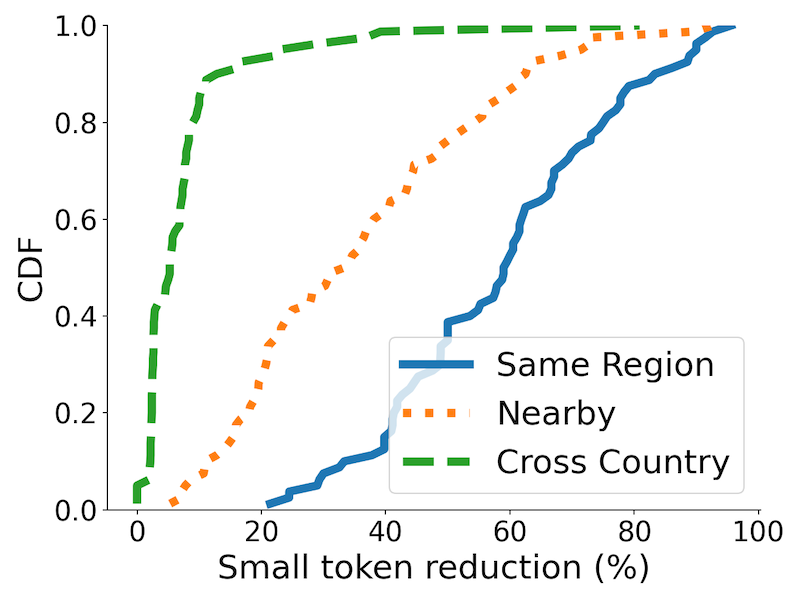}\label{subfig:tokens-vllm}}

    \caption{Evaluation of \sys{} on MTBench implemented in vLLM and deployed in AWS. We evaluated 3 deployments of the controller/worker: both in the same region, across the two nearby us-east-* regions, and across the country in \texttt{us-east-1} and \texttt{us-west-2}. Each deployment is compared to speculative decoding on a single machine. \ref{subfig:time-vllm} shows the reduction in generating responses and \ref{subfig:tokens-vllm} shows the reduction in speculative tokens decoded by the controller. Using two regions on the same coast, ~20ms RTT, allowed significant token reduction while slightly improving latency.}
    \label{fig:vllm-results}
\end{figure}



We also implement the \sys{} controller and worker in vLLM~\cite{vllm}. We deployed the modified controller in the AWS \texttt{us-east-1} region. Three workers were deployed in \texttt{us-east-1}, \texttt{us-east-2}, and \texttt{us-west-2}. Each used a \texttt{g6e.8xlarge} which has a 48GB NVIDIA L40S GPU. This implementation uses a branch factor, $b$, of 1 (no branching). The RTT between controller and worker for the three setups is approximately 10ms, 15ms, and 70ms. Unlike the simulation, this experiment runs real target/draft models, Llama-3.1-8B/Llama-3.2-1B, respectively~\cite{grattafiori2024llama3herdmodels}. On this hardware steps of the target model take $\sim$23.4ms, and the draft models takes $\sim$7.5ms.

We again generate 100 token responses to the MTBench dataset~\cite{mtbench}. With the models running directly, we no longer need to rely on the the i.i.d. assumption from simulation. We observed a 75\% rate of matches between tokens predicted by the draft and target model on this workload. We quantified deviation from the simulator’s i.i.d. assumption: in $k=2$ batches the first token matches 80\% of the time, while the second matches 83\% conditioned on the first match. The difference is statistically significant but small.

Our experiment results in Fig~\ref{fig:vllm-results} confirm that \sys{} can still reduce draft token decoding on the controller while avoiding large increases in latency with a real world workload. As we expected from simulation, latency between controller and worker of around 15ms can see significant shifting of draft model load. However, larger cross country latencies have diminished improvements. \sys{} is designed to fall back to the same behavior as standard speculative decoding at high RTTs. The results of our cross country experiments confirm these diminishing returns, with some increase in generation time when high RTTs prevent the benefits of offloading. Outliers start to show up with high RTTs as well, which can be caused by unexpected latency spikes. For these reasons it's best to disable \sys{} when latencies are too high to allow useful offloading. The results of our cloud deployment give us confidence that the properties of a real world deployment not present in our simulator such as a KV cache and real output distributions do not significantly impact our design.



%% file: related_work.tex
\section{Related Work}
\label{related_work}

\subsection{LLM Measurements}
Prior work evaluating LLM serving systems have used real world workloads for request/response tokens such as ShareGPT~\cite{sharegpt} and Alpaca~\cite{alpaca}. BurstGPT~\cite{wang2025burstgptrealworldworkloaddataset} introduced a real world workload measured from a cloud provider in a single region that demonstrated varying requests patterns in conversation vs. API traffic. Our measurements in~\S\ref{sec:motivation} spanning multiple data centers across the globe show that latency patterns vary significantly by region. Some regions exhibit much stronger diurnal patterns, and others have greater tail latency.

\subsection{Speculative decoding}
Many extensions to speculative decoding have been proposed, a survey of different techniques highlights advantages of each~\cite{survey-speculative}. One approach gaining popularity is EAGLE~\cite{li2025eaglespeculativesamplingrequires} which predicts the feature layer rather than tokens, and EAGLE-2~\cite{li2024eagle2fasterinferencelanguage} which incorporates dynamic draft trees. Other work on speculative decoding has also proposed various implementations of draft trees~\cite{spec_infer, opt-tree}. Other speculative decoding work asynchronously runs the draft and target models by splitting GPU resources on a single machine~\cite{zhang2025swiftspecultralowlatencyllm}. Still others have applied speculative generation techniques to consumer devices using parallelization and offloading~\cite{svirschevski2024specexec, dovetail}. With some modification, \sys{} can benefit from optimizations proposed in these related works to improve the accuracy of the worker. This would in turn reduce the frequency of synchronization steps and amplify the benefits of \sys{}.

\subsection{LLM serving systems}
Many LLM-specific serving systems have also been proposed. Orca~\cite{orca} introduced iteration level scheduling to improve utilization of batched requests. Other scheduling improvements including Sarathi-Serve\cite{agrawal2024taming} and DistServe~\cite{dist-serve} focus on scheduling the unique requirements of prefill and decoding stages. vLLM~\cite{vllm} introduced PagedAttenion to manage memory of the KV cache inspired by virtual memory. FlashAttention~\cite{dao2022flashattentionfastmemoryefficientexact} also bases its optimization on memory management by reducing access to the slowest memory. TensorRT-LLM~\cite{tensorRT} optimizes model inference for NVIDIA GPUs. Predicting the relative length of responses has helped systems reduce head of line blocking in LLM request scheduling~\cite{fu2024efficient}. Many of these systems support using speculative decoding, and could be modified to support~\sys{}.

%% file: conclusion.tex
\section{Conclusion}
\label{conclusion}

We present \sys{}, a speculative decoding strategy that operates over a wide area network. We use judicious redundancy and entropy based heuristics to avoid long stalls during synchronization across the WAN. Our proposal to distribute inference across multiple regions is supported by our measurement study, demonstrating varying latency characteristics of different data center regions. In future work we plan to extend this novel speculative decoding strategy to a distributed inference serving system by considering batching and priorities of various request types.